\documentclass[preprint,nofootinbib,preprintnumbers]{revtex4}

\usepackage[dvips]{graphicx}
\usepackage{amsmath,amsthm,amssymb}

\begin{document}

\title{The Early-Time Evolution of the Cosmological Perturbations\\ in $f(R)$ Gravity}

\author{Je-An Gu$^{1}$}
\email{jagu@ntu.edu.tw}
\author{Tse-Chun Wang$^{2}$}
\email{r98222067@ntu.edu.tw}
\author{Yen-Ting Wu$^{1,2}$}
\email{r97222061@ntu.edu.tw}
\author{Pisin Chen$^{1,2,3,4}$}
\email{pisinchen@phys.ntu.edu.tw}
\author{W-Y.~Pauchy Hwang$^{2,3,5,6}$}
\email{wyhwang@phys.ntu.edu.tw}

\affiliation{ \vspace{0.66em}
${}^1$Leung Center for Cosmology and Particle Astrophysics,\\
${}^2$Department of Physics, and ${}^3$Graduate Institute of Astrophysics,\\
National Taiwan University, Taipei 10617, Taiwan, R.O.C.\\
${}^4$Kavli Institute for Particle Astrophysics and Cosmology, SLAC National Accelerator Laboratory, Menlo Park, CA 94025, U.S.A.\\
${}^5$Asia Pacific Organization for Cosmology and Particle Astrophysics, and\\
${}^6$Center for Theoretical Physics, National Taiwan University, Taipei 10617, Taiwan, R.O.C.%
}

\date{\today}

\begin{abstract}
We investigate the evolution of the linear cosmological
perturbations in $f(R)$ gravity, an alternative to dark energy
for explaining the late-time cosmic acceleration. We
numerically calculate the early-time evolution with an
approximation we contrive to solve a problem that commonly
appears when one solves the full evolution equations. With the
approximate evolution equations we can fairly assess the effect
of the gravity modification on the early-time evolution,
thereby examining the validity of the general-relativity (GR)
approximation that is widely used for the early universe. In
particular, we compare the CMB photon
density perturbation 
and the matter density perturbation 
obtained respectively by our approximation and the conventional
GR approximation. We find that the effect of the gravity
modification at early times in $f(R)$ gravity may not be
negligible.
We conclude that to be self-consistent, in the $f(R)$ theory
one should employ the approximation presented in this paper
instead of that of GR in the treatment of the early-time
evolution.
\end{abstract}


\maketitle

\section{Introduction} \label{sec:introduction}

The accelerating expansion of the present universe can be
explained by an energy source of anti-gravity, generally termed
dark energy, or alternatively by the large-scale, low-energy
modification of the gravity theory. In this paper we focus on
the $f(R)$ theory of modified gravity
(for a review, see \cite{DeFelice:2010aj,Nojiri:2010wj})
with the gravity action
\begin{equation}\label{actionfR}
S_{g} = \frac{1}{16\pi G_N} \int d^4x \sqrt{-g} [R+f(R)] .
\end{equation}
In this theory the deviation
from general relativity (GR) is represented by a function of
the Ricci scalar, $f(R)$, within the gravity
action.\footnote{We consider the metric formalism of $f(R)$ gravity
and use the natural units where $c=\hbar=1$.}


The gravitational field equations obtained from the above
action are
\begin{equation}\label{fRfieldeq}
\begin{aligned}
(1+f_R)R_{\mu\nu} - \frac{1}{2}(R+f)g_{\mu\nu}
+ (g_{\mu\nu}\square-\nabla_\mu\nabla_\nu)f_R
= 8\pi G_N T_{\mu\nu} \,,
\end{aligned}
\end{equation}
where $\square$ is the d'Alembertian. We use the notation,
$f_R \equiv df/dR$ and $f_{RR} \equiv d^2f/dR^2$, in this paper.

For the background expansion of the universe we consider a
homogeneous and isotropic space-time described by the flat
Robertson-Walker metric:
\begin{equation}\label{FRLW}
ds^2 = a^{2}(\tau)\left\{-d\tau^{2}+\vec{x}^2\right\} ,
\end{equation}
where $a$ is the scale factor and $\tau$ is the conformal time. With
this metric the above gravitational field equations lead to
\begin{equation}\label{eq:Friedmann-eq1}
H^2+H^2f_R+\frac{f}{6}+\frac{H}{a}\dot{f}_R-\frac{1}{6}Rf_R=\frac{8\pi G_N}{3}\rho \,,
\end{equation}
\begin{equation}\label{eq:Friedmann-eq2}
\begin{aligned}
\left(H^2-\frac{R}{3}\right)+f_R\left(\frac{R}{6}+H^2\right)-\frac{f}{2}-\frac{\ddot{f}_R}{a^2}-\frac{H}{a}\dot{f}_R=8\pi G_N P \,,
\end{aligned}
\end{equation}
where the overhead dot denotes the derivative w.r.t.\ the conformal time
$\tau$, the Hubble expansion rate $H = \dot{a}/a^2$,
and $\rho$ and $P$ are the average energy density and the
average pressure of the universe, for which we will consider
matter and radiation.

For a given expansion history $a(\tau)$, as well as given
$\rho(a)$ and $P(a)$, Eq.\ (\ref{eq:Friedmann-eq1}) becomes a
second-order differential equation of $f(\tau)$ or $f(R)$. The
functions $f(R)$ that satisfy this equation can generate the
required expansion history. On the other hand, the dark energy
models can also generate the required expansion history by
choosing an appropriate dark energy density
$\rho_\textsc{de}(a)$ and pressure $P_\textsc{de}(a)$.
Consequently, measurements of cosmic expansion alone cannot
distinguish $f(R)$ gravity from dark energy, and additional
independent measurements such as the cosmic structures are
indispensable.

For the cosmic structure formation in $f(R)$ gravity, people
studied the evolution of the cosmological perturbations
\cite{Carroll:2006jn,Song:2006ej,Bean:2006up}.
While the evolution at late times has been widely studied
\cite{Tsujikawa:2007gd,delaCruzDombriz:2008cp,Tsujikawa:2009ku},
the evolution at early times is typically treated with a simple
approximation,
the GR approximation, where the deviation from
GR is ignored. 
(For a treatment different from the GR approximation for the
early times, see \cite{Bean:2006up} where the evolution from
the early times to the present is studied.)

In this paper we take into account and carefully investigate
the effect of the gravity modification in $f(R)$ gravity on the
early-time evolution of the linear perturbations. When
numerically solving the full evolution equations in $f(R)$
gravity for the early times, one is usually confronted with a
tight-coupling issue. To solve this issue we contrive a better
approximation, with which we can fairly assess the effect of
the gravity modification at early times on the evolution. With
those at hand, we then examine the validity of the conventional
GR approximation. In particular, we will compare the density
perturbations of  the CMB photons and matter 
obtained respectively by our approximation and the conventional
GR approximation. We will show that the effect of the gravity
modification at early times in $f(R)$ gravity may not be
negligible. Accordingly, for the early-time evolution of the
perturbations in $f(R)$ gravity, the GR approximation is
problematic, and a better treatment is necessary.

\section{Cosmological Perturbations in $f(R)$ Gravity} \label{sec:cosmo-perturb}
For the cosmological perturbations in the early universe we
analyze the evolution equations of the linear perturbations in
the Fourier space and in the synchronous gauge
\cite{Ma:1995ey}. For the metric perturbations we consider the
scalar modes, $h(\vec{k},\tau)$ and $\eta(\vec{k},\tau)$,
defined by the line element:
\begin{equation}\label{02}
\begin{aligned}
ds^{2}=a^{2}(\tau)\left\{-d\tau^{2}
+\left[\delta_{ij}+h_{ij}(\vec{x},\tau)\right]dx^{i}dx^{j}\right\} ,
\end{aligned}
\end{equation}
and the Fourier integral:
\begin{equation}\label{03}
\begin{aligned}
h_{ij}(\vec{x},\tau)=\int d^{3}k e^{i\vec{k}\cdot\vec{x}} \left[
{\hat{k_{i}}\hat{k_{j}}h(\vec{k},\tau) +\left(\hat{k_{i}}\hat{k_{j}}-\frac{1}{3}\delta_{ij}\right)6\eta(\vec{k},\tau)} \right] ,
\end{aligned}
\end{equation}
where $\vec{k}=k\hat{k}$ and $k$ is the comoving wave number.
With regard to the energy part, we consider the stress-energy
perturbations of cold dark matter (CDM), baryons, photons and
massless neutrinos. 
For each of the particle species,
\begin{eqnarray}
\delta(\vec{k},\tau) &\equiv& \delta\rho(\vec{k},\tau)/\rho \,,\\
\theta(\vec{k},\tau) &\equiv& i k^{j}\delta T^0{}_{\!j}(\vec{k},\tau)/(\rho+P) \,,\\
\sigma(\vec{k},\tau) &\equiv& -(\hat{k_{i}}\hat{k_{j}}-\frac{1}{3}\delta_{ij})\Sigma^i{}_{\!j}(\vec{k},\tau)/(\rho+P) \,,
\end{eqnarray}
where $\Sigma^i{}_{\!j}\equiv
T^i{}_{\!j}-\delta^i{}_{\!j}T^k{}_{\!k}/3$.

For organizing the evolution equations of the above perturbed
quantities, we introduce two new dynamical variables:
\begin{eqnarray}
q &\equiv& \dot{h}+6\dot{\eta} \,,\label{frfield6}\\
\chi &\equiv& f_{RR}\delta R_N \,,\label{}
\end{eqnarray}
where $\delta R_N$ is defined as the perturbation of the Ricci
scalar in the conformal Newtonian gauge, and its relation to
the metric perturbations in the synchronous gauge is:
\begin{equation} \label{RN}
\delta R_N = -\frac{6}{a^2}\ddot{\eta} - \frac{18H}{a}\dot{\eta}
- \frac{4k^2}{a^2}\eta
+ \frac{1}{a^2}\dot{q}
+ 3 \left( \frac{\ddot{H}}{ak^2} + \frac{3H\dot{H}}{k^2} + \frac{H}{a}
\right) q \, .
\end{equation}

\subsection{Evolution Equations} \label{sec:evolution-eqns}
Since the evolution equations of the stress-energy
perturbations in $f(R)$ gravity are given by the Boltzmann
equations with the same form as those in GR \cite{Ma:1995ey},%
\footnote{The Boltzmann equations describe the microscopic
physics and therefore the form of the equations is independent
of the gravity theories.}
here we will simply present the evolution equations of the
metric perturbations, $q$, $\eta$, and $\chi$. (The information
about $h$ can be derived from that about $q$ and $\eta$.)

The $i$--$j$ component ($i \neq j$) of the gravitational field
equations in Eq.\ (\ref{fRfieldeq}) gives
\begin{equation}\label{qdot}
\begin{aligned}
\dot{q} = -2aHq+2k^2\eta-\frac{2k^2\chi}{1+f_R}-\frac{12\pi G_Na^2}{1+f_R}\sum_a\rho_a\sigma_a\left(1+w_a\right) .
\end{aligned}
\end{equation}
The linear combinations of the $0$--$0$ and $0$--$j$
components give
\begin{equation}\label{etadot}
\begin{aligned}
\dot{\eta} =
& \left(\frac{a^2H^2}{k^2\dot{f}_R}\right)
\left\{ \left[ \frac{k^2\left(1+f_R\right)}{3aH}
-\frac{\dot{H}\dot{f}_R}{2aH^2} \right] q
-\frac{2k^4\left(1+f_R\right)}{3a^2H^2}\eta \right.
+\left(\frac{k^4}{3a^2H^2}-\frac{k^2\dot{H}}{aH^2}\right)\chi \\
& \left. \hspace{5em}
-\frac{8\pi G_Na}{H}\sum_a\rho_a\theta_a\left(1+w_a\right)
-\frac{8\pi G_Nk^2}{3H^2}\sum_a\rho_a\delta_a
\right\} ,
\end{aligned}
\end{equation}
\begin{equation}\label{chidot}
\begin{aligned}
\dot{\chi} =
& -\frac{aH}{k^2}\left[\frac{\dot{f}_R}{2}+\frac{\dot{H}}{H}f_R
-\frac{4\pi G_Na}{H}\rho_\textrm{eff}\left(1+w_\textrm{eff}\right)\right]q
+\dot{f}_R\eta +2\left(1+f_R\right)\dot{\eta}
+\left(aH-\frac{\dot{f}_R}{1+f_R}\right)\chi\\
& -\frac{8\pi G_Na^2}{k^2}\sum_a\rho_a\theta_a\left(1+w_a\right)
-\frac{12\pi G_Na^2\dot{f}_R}{k^2\left(1+f_R\right)}\sum_a\rho_a\sigma_a\left(1+w_a\right) .
\end{aligned}
\end{equation}
In these three evolution equations 
the subscript $a$ runs over the particle species including CDM, baryons,
photons and massless neutrinos. In Eqs.\ (\ref{etadot}) and (\ref{chidot})
the effective energy density $\rho_\textrm{eff}$ and the
effective equation of state $w_\textrm{eff}$ are defined as \cite{Amendola:2006we}
\begin{eqnarray}
\rho_\textrm{eff}&\equiv&\frac{1}{8\pi G_N}\left(\frac{1}{2}Rf_R-3H^2f_R-\frac{f}{2}-\frac{3H}{a}\dot{f}_R\right),\label{}\\
w_\textrm{eff}&\equiv&\frac{P_\textrm{eff}}{\rho_\textrm{eff}}=
-\frac{1}{3}-\frac{2}{3}\left(\frac{-\frac{1}{2a^2}\ddot{f}_R-\frac{1}{6}f+H^2f_R}{-\frac{H}{a}\dot{f}_R-H^2f_R+\frac{1}{6}R f_R-\frac{1}{6}f}\right).\label{}
\end{eqnarray}
They characterize the effects of the modification of gravity at
the background expansion level.

\subsection{A Problem of Solving the Full Evolution Equations}
\label{sec:numerical-problem}
One is confronted with a problem when numerically
solving the above evolution equations for the early universe,
particularly Eq.\ (\ref{etadot}). Here we elucidate the problem.

We reorganize Eq.\ (\ref{etadot}) as follows.
\begin{equation}\label{etadot-reorganize}
\begin{aligned}
\dot{\eta} =
& \left(\frac{a^2H^2}{k^2}\right)
\left\{
\left[
\left(\frac{k^2}{3aH}\right)q
-\frac{2k^4}{3a^2H^2}\eta
-\frac{8\pi G_Na}{H}\sum_a\rho_a\theta_a\left(1+w_a\right)
-\frac{8\pi G_Nk^2}{3H^2}\sum_a\rho_a\delta_a
\right] \right. \\
& \left. \left.
+\left[
 \left(\frac{k^2}{3aH}f_R-\frac{\dot{H}}{2aH^2}\dot{f}_R\right)q
-\frac{2k^4}{3a^2H^2}f_R\eta
+\left(\frac{k^4}{3a^2H^2}-\frac{k^2\dot{H}}{aH^2}\right)\chi
\right]
\right\} \right/ \dot{f}_R \, .
\end{aligned}
\end{equation}
This equation can be read as
\begin{equation}\label{etadot-simple}
\dot{\eta} =
\left[\textrm{GR-terms}+f\textrm{-terms}\right] / \dot{f}_R \, ,
\end{equation}
where the ``$f$-terms'' denote the terms proportional to the
derivatives of $f$ (including $\chi$), and the ``GR-terms'' are
the other terms that also appear in the GR equations.
When the $f(R)$ theory is very close to GR (e.g., at early
times), $\dot{f}_R$ and the $f$-terms are much smaller than the
GR-terms. In this case, to correctly obtain $\dot{\eta}$ via
the above equation,
the summation of the GR-terms should be as small as $\dot{f}_R \dot{\eta}$. However, the error in calculating each GR term in the perturbation theory,
i.e., due to ignoring high-order perturbations, can be much larger than
$\dot{f}_R \dot{\eta}$, thereby making the calculation
of $\dot{\eta}$ in Eq.\ (\ref{etadot}) incorrect.

This problem is analogous to the issue caused by the tight
coupling between photons and baryons (before the decoupling
around $a \sim 10^{-3}$) that renders the evolution equations
of the perturbations in the standard cosmology difficult to
solve. A simple but rough solution is to invoke the
approximation where baryons and photons behave like a single
coupled fluid with $\dot{\theta}_{\gamma} = \dot{\theta}_b$.
Instead, cosmologists invoke a more accurate approximation,
termed ``tight-coupling approximation'', to account for the
slip between the photon and baryon fluids \cite{Ma:1995ey}.

In our case for $f(R)$ gravity at early times, the largeness of
the factor $1/\dot{f}_R$ leads to the tight coupling among the
GR-terms, which makes the calculation of $\dot{\eta}$ in Eq.\
(\ref{etadot}) incorrect. Similarly, a simple but rough
solution is to use the GR approximation in Eq.\ (\ref{etadot})
such that the summation of the GR-terms vanishes and the
$f$-terms and $\dot{f}_R \dot{\eta}$ are ignored. Nevertheless,
to have better accuracy we contrive an approximation to account
for the modification of gravity in $f(R)$ gravity.

The tightness of the coupling can be characterized by
$aH/\dot{f}_R$ (for the super-horizon modes) or by $(k^2/aH) /
\dot{f}_R$ (for the sub-horizon modes), where the largeness of
which suggests the tightness. For demonstration we will show in
Sec.\ \ref{sec:results} the evolution of $aH/\dot{f}_R$ in a
$f(R)$ model, where 
$aH/\dot{f}_R \sim 10^{11}$ when $a \sim 10^{-3}$ and $\sim
10^{7}$ when $a \sim 10^{-2}$.


We know that the CMB observational results are consistent with
GR with high precision. Therefore the allowed deviation from GR
at early times must be small.
Accordingly, in the viable $f(R)$ models of the late-time
cosmic acceleration, $f_R$ and its derivatives should be tiny
at early times, which leads to the tight-coupling issue.

\subsection{``Tight-Coupling'' Approximation}
\label{sec:our-approx}
For the early-time evolution we construct a new approximation
to solve the problem discussed above. In dealing with Eq.\
(\ref{etadot}) we decompose $\eta$ into two parts,
\begin{equation} \label{etaorder}
\eta=\eta^{(0)}+\eta^{(1)} ,
\end{equation}
where $\eta^{(0)}$ and $\eta^{(1)}$ are designed to be
comparable respectively to the GR-terms and $f$-terms
normalized by $H^2$, so that Eq.\ (\ref{etadot}) can also be
divided into two parts which respectively lead to the evolution
equations of $\eta^{(0)}$ and $\eta^{(1)}$ with no
tight-coupling issue.
For this purpose, we set
\begin{equation}\label{etadot0}
\begin{aligned}
\dot{\eta}^{(0)}\equiv\frac{4\pi G_Na^2}{k^2}\sum_a\rho_a\theta_a\left(1+w_a\right)-\frac{2\pi G_Na^2}{\left(1+f_R\right)k^2}\rho_\textrm{eff}\left(1+w_\textrm{eff}\right)q \,,
\end{aligned}
\end{equation}
and then derive the evolution equation of $\eta^{(1)}$:
\begin{equation}\label{etadot1}
\begin{aligned}
\dot{\eta}^{(1)}=&\frac{1}{1+f_R}\left[\frac{1}{2}\dot{\chi}+\left(\frac{aH}{4k^2}\dot{f_R}+\frac{a\dot{H}}{2k^2}f_R\right)q-f_R\dot{\eta}^{(0)}
-\frac{1}{2}\dot{f_R}\left(\eta^{(0)}+\eta^{(1)}\right)\right.\\&
\left.-\frac{1}{2}\left(aH-\frac{\dot{f_R}}{1+f_R}\right)\chi+\frac{6\pi G_Na^2\dot{f_R}}{k^2(1+f_R)}\sum_a\rho_a\sigma_a\left(1+w_a\right)\right] .
\end{aligned}
\end{equation}

Solving Eq.\ (\ref{etadot1}) requires the information about
$\chi$ and $\dot{\chi}$. Instead of using Eq.\ (\ref{chidot}),
to solve the problem we invoke the following approximation for
$\chi$:
\begin{equation}\label{chi0}
\chi \equiv f_{RR} \delta R_N
\approx \chi_{(\textrm{approx})} \equiv -8\pi G_Nf_{RR}\delta T_N
= 8\pi G_N f_{RR}\sum_{a}\rho_{a}(1-3w_a)\delta_{N,a} \,,
\end{equation}
that is, $\delta R_N 
\approx -8\pi G_N\delta T_N$,
where  $\delta T_N$ is the perturbation of the trace of the stress-energy tensor in the conformal Newtonian gauge, $\delta T_N \equiv (\delta T^{\mu}{}_{\!\mu})_N $,
and $\delta_{N,a}$ is the density perturbation of the $a$-th fluid in the conformal Newtonian gauge \cite{Ma:1995ey}.

With regard to $\dot{\chi}$, we derive its relation to other
perturbed quantities from the time derivative of Eq.\
({\ref{chi0}}):
\begin{equation}\label{chi0dot}
\begin{aligned}
\dot{\chi} 
\approx \dot{\chi}_{(\textrm{approx})} \equiv\
& 8\pi G_N \sum_{a}\rho_{a}\left[\dot{f}_{RR}-3aH(1+w_a)f_{RR}\right]
(1-3w_a)\delta_{N,a} \\
& +8\pi G_N f_{RR}\sum_{a}\rho_{a}(1-3w_a)\dot{\delta}_{N,a}^{(0)} \,,
\end{aligned}
\end{equation}
where
\begin{equation}\label{}
\dot{\delta}_{N,a}^{(0)}\equiv(1+w_a)\left(-\theta_a-{q\over 2}+3\dot{\eta}^{(0)} - {3aH\over 2k^2}\dot{q} - {3a^2H^2+3a\dot{H}\over 2k^2}q \right) ,
\end{equation}
That is, we neglect $\dot{\eta}^{(1)}$ when calculating
$\dot{\delta}_{N,a}$ in the above $\dot{\chi}$ relation, as the
second approximation. This approximation and that in Eq.\
(\ref{chi0}) are the two approximations we make in our
treatment of the early-time evolution.

To calculate the early-time evolution of the perturbations in
$f(R)$ gravity, we solve the set of the coupled evolution
equations and relations including Eqs.\ (\ref{qdot}),
(\ref{etaorder}), (\ref{etadot0}), (\ref{etadot1}),
(\ref{chi0}), (\ref{chi0dot}), and the Boltzmann equations. No
tight coupling appears in this set of equations.

\subsection{GR Approximation vs.\ Tight-Coupling Approximation}
\label{sec:GR-approx}
Conventionally people take the GR approximation
\cite{Tsujikawa:2007gd,Tsujikawa:2009ku}, where the early
universe is described by the $\Lambda$CDM model, to solve the
early-time evolution equations of the perturbations in $f(R)$
gravity, thereby giving an initial condition for the late-time
evolution equations under the matter-domination approximation
(and maybe other approximations). In many cases people solve
the late-time approximate evolution equations from an initial
time between $a=0.01$ and $a=0.03$. That is, it is widely
believed that the GR approximation is valid to a high precision
at least before $a=0.01$ for most viable $f(R)$ models.

In the conventional method the effects of the modification of
gravity at early times are neglected and therefore can hardly
be assessed. On the contrary, our approximation takes into
account the effect of the gravity modification in $f(R)$
gravity. Our approximate equations in Sec.\
\ref{sec:our-approx} go back to the evolution equations in GR
when $f_R$ and $f_{RR}$ go to zero, i.e., when the effects of
the gravity modification are eliminated. Therefore, the GR
approximation is a limiting case of our approximation and also
a rougher approximation than ours. With our approximation we
can assess the effect of the gravity modification on the
early-time evolution, thereby examining the validity of the GR
approximation.


\section{Results} \label{sec:results}
We compare the early-time evolution of the cosmological
perturbations obtained respectively by our approximation and
the GR approximation.
We modify the CMBFAST code \cite{Seljak:1996is} to numerically
solve our approximate early-time evolution equations of the
cosmological perturbations in $f(R)$ gravity, while we use
CMBFAST to obtain the early-time evolution under the GR
approximation.

For the purpose of demonstration, we consider a designer $f(R)$
model \cite{Pogosian:2007sw} 
with $w_\textrm{eff}=-1$ and the initial condition:
$f_{R}(a_i)=-1.3923 \times 10^{-39}$ at $a_i = 10^{-8}$.
This model is 
consistent with the observational results about the cosmic structures \cite{Lin:2010hk}.%
\footnote{A designer $f(R)$ model with the effective equation
of state $w_\textrm{eff}$ gives the same expansion history as
that of a dark energy model with $w_\textsc{de} =
w_\textrm{eff}$. We invoke the code developed by Wei-Ting Lin
to numerically calculate $f(R)$ and its derivatives for given
$w_\textrm{eff}$, $f_R(a_i)$, and the values of other
cosmological parameters.}
With regard to the other cosmological parameters, we use the
values suggested by the Seven-Year Wilkinson Microwave
Anisotropy Probe (WMAP7) observations \cite{Komatsu:2010fb}:
The effective number of neutrino species $N_\textrm{eff}=4.34$,
the mass fraction of helium $Y_\textrm{He}=3.26$, the Hubble
constant $H_{0}=73.8\,$km/s/Mpc, the baryon density fraction
$\Omega_{b0}=0.0455$, the cold dark matter $\Omega_{c0}=0.226$,
the effective dark energy $\Omega_\textrm{eff0}=0.728$, and the
matter-radiation equality time $z_\textrm{eq}=4828$.

\begin{figure}
\includegraphics{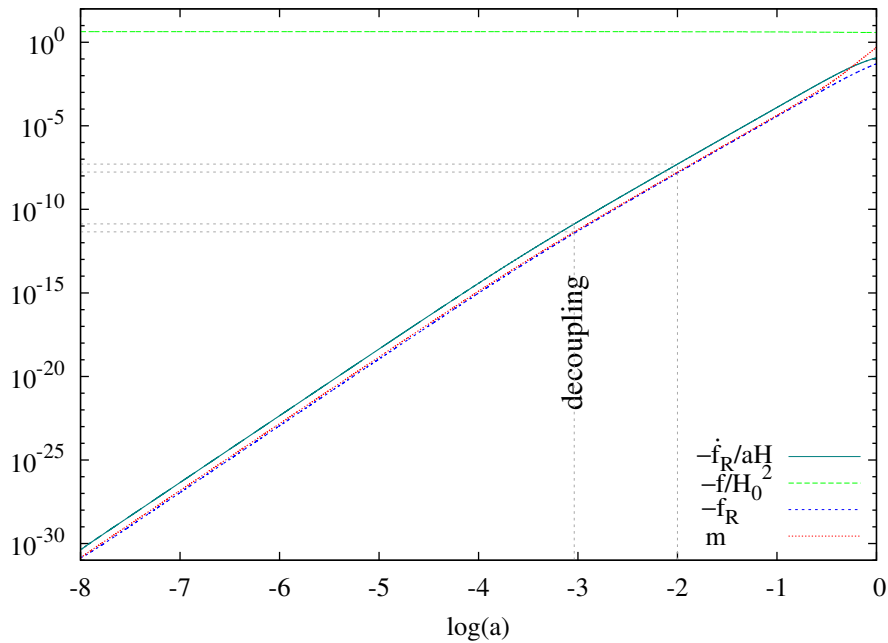}
\caption{The evolution of $-\dot{f}_R/aH$, $-f/H_0^2\,$, $-f_R$ and $m \equiv Rf_{RR}/(1+f_R)$ for the designer $f(R)$ model with $w_\textrm{eff}=-1$ and the initial condition: $f_R(a_i) = -1.3923 \times 10^{-39}$ at $a_i = 10^{-8}$.}
\label{design}
\end{figure}

Figure \ref{design} shows the evolution of several $f$-related
quantities for the designer $f(R)$ model under consideration,
including $-f/H_0^2$ and the derivatives: $-\dot{f}_R/aH$
(introduced in Sec.\ \ref{sec:numerical-problem}), $-f_R$ and
$m \equiv Rf_{RR}/(1+f_R)$ \cite{Amendola:2006we}. The quantity
$m$ is conventionally used to characterize the deviation from
GR
\cite{Tsujikawa:2007gd,delaCruzDombriz:2008cp,Tsujikawa:2009ku}.
In this model the derivatives of $f$ grow with time from tiny
values at early times to the order of unity at present, and
accordingly $f$ is nearly a constant at early times and
slightly changes in the recent epoch around the value
$-2\Lambda$.

\begin{figure}
\includegraphics{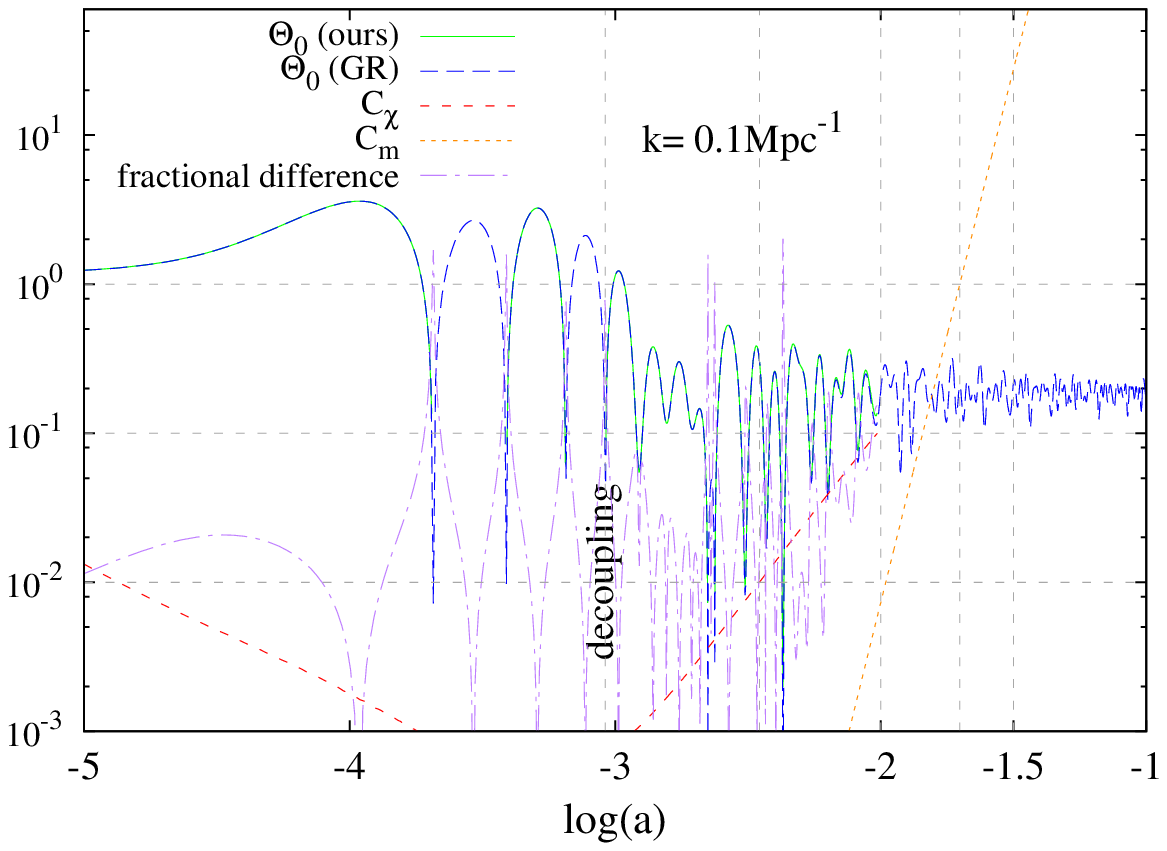}
\includegraphics{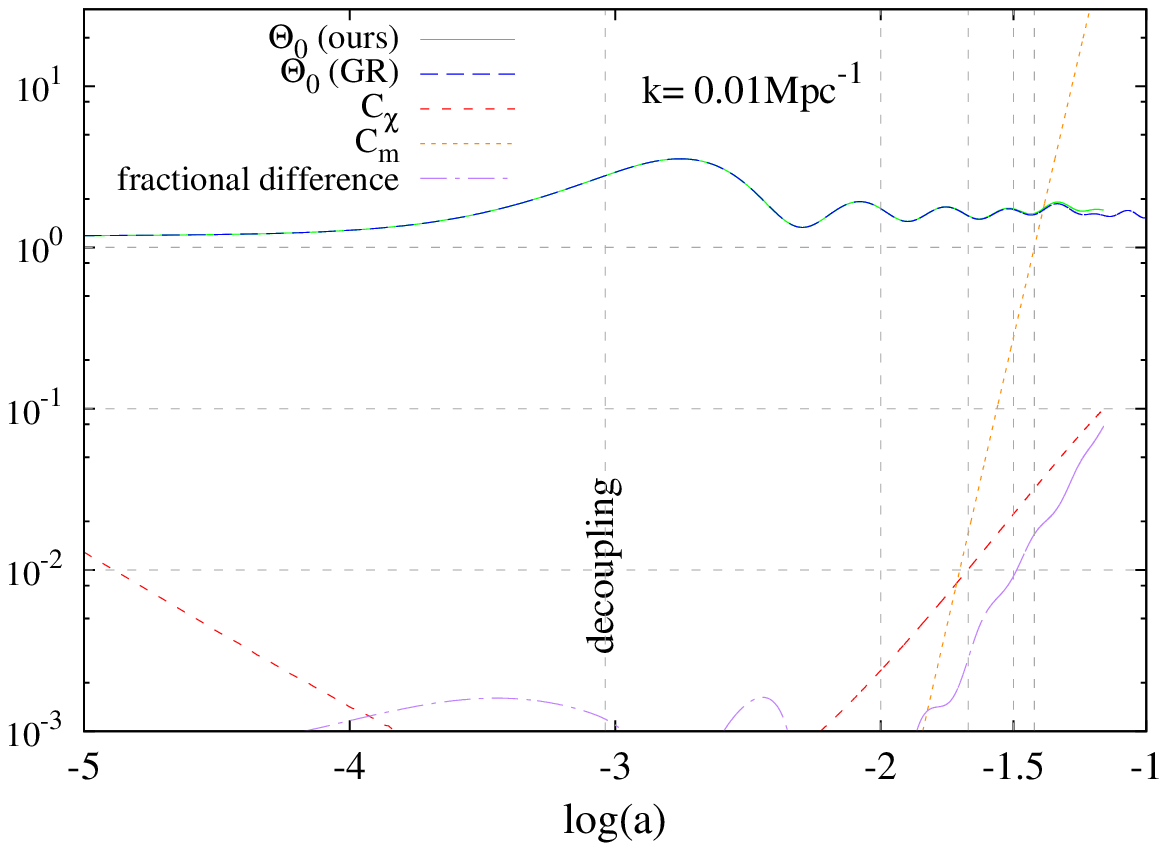}
\caption{The comparison of the CMB photon density perturbations obtained respectively by our approximation $\Theta_0(\textrm{ours})$ and the GR approximation $\Theta_0(\textrm{GR})$. The upper panel is for the case where $k=0.1\,$Mpc$^{-1}$, and the lower panel for $k=0.01\,$Mpc$^{-1}$.
We present the evolution of the fractional difference in $\Theta_0$ between these two approximations, as well as the fractional difference $c_{\chi}$ between $\chi_\textrm{(approx)}$ and $\chi$ as an indicator of the validity of our approximation,
and $c_m \equiv (aH/k)^2 m$ (where $m \equiv R f_{RR}/(1+f_R)$)
that is related to the validity of the sub-horizon approximation.
}
\label{photon}
\end{figure}

\begin{figure}
\includegraphics{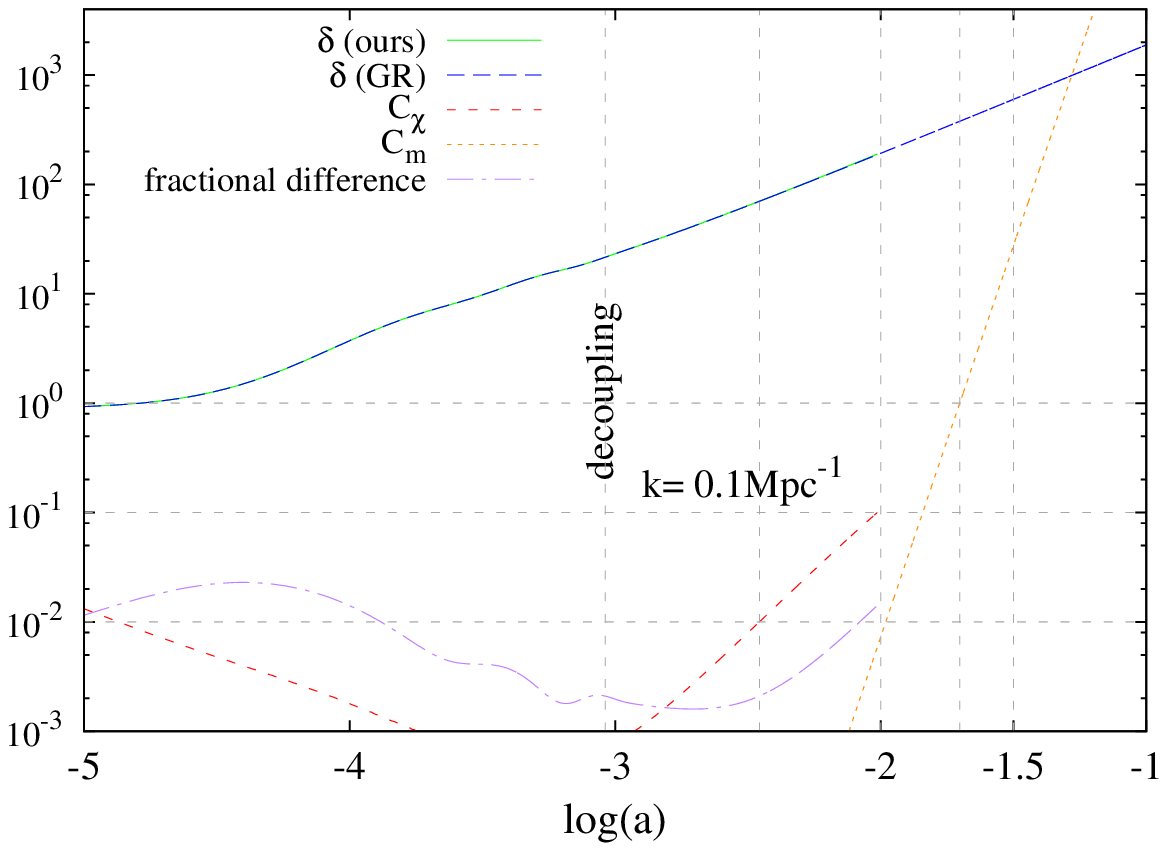}
\includegraphics{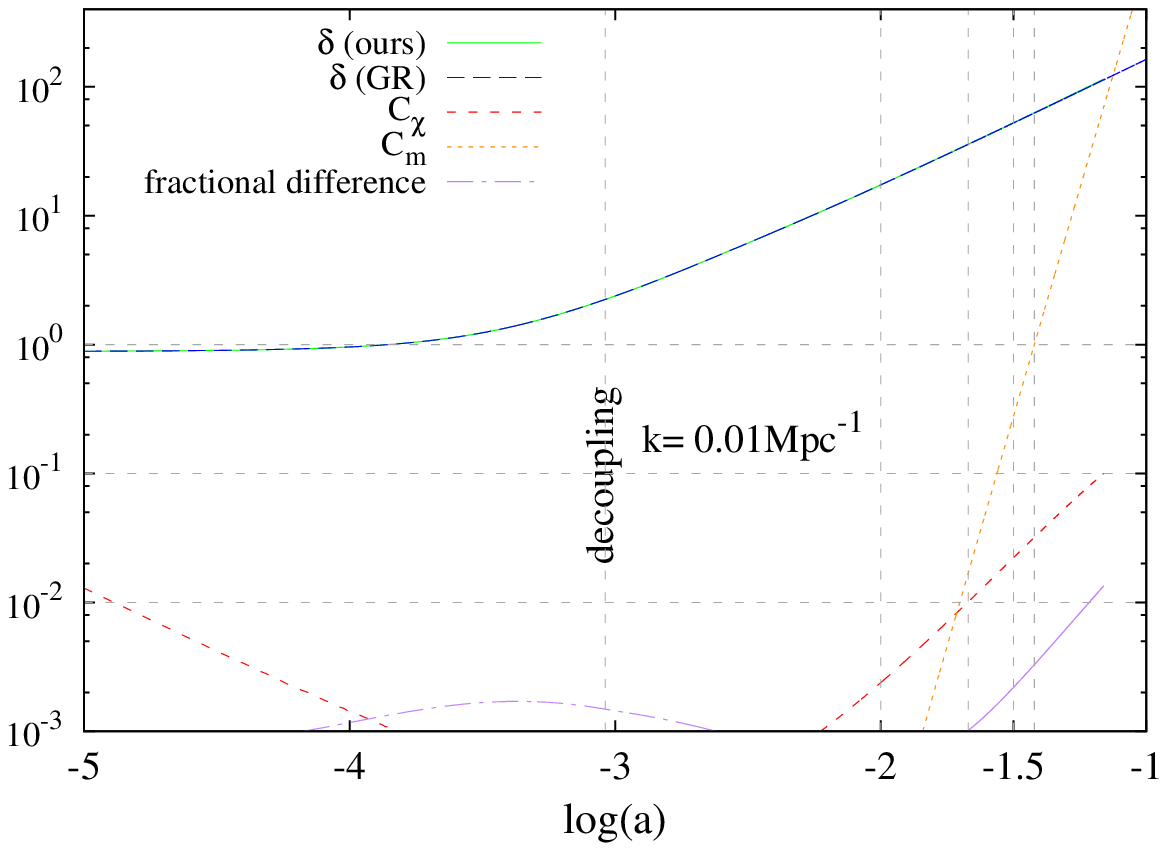}
\caption{The comparison of the matter density perturbations obtained respectively by our approximation $\delta(\textrm{ours})$ and the GR approximation $\delta(\textrm{GR})$. The upper and the lower figures are respectively for $k=0.1\,$Mpc$^{-1}$ and $k=0.01\,$Mpc$^{-1}$. We present the evolution of the fractional difference in $\delta$ between these two approximations, as well as the evolution of $c_{\chi}$ and $c_m$.
\label{deltam}
}
\end{figure}

We present the evolution of two Fourier modes, 
$k=0.1\,$Mpc$^{-1}$ and $k=0.01\,$Mpc$^{-1}$. In Fig.\
\ref{photon} we present the CMB photon density perturbation
$\Theta_0$ in the synchronous gauge and its fractional
difference between our approximation and the GR approximation,
$\left|\Theta_0(\textrm{ours})-\Theta_0(\textrm{GR})\right| /
\left[
\left|\Theta_0(\textrm{ours})\right|+\left|\Theta_0(\textrm{GR})\right|
\right]$. In Fig.\ \ref{deltam} we present the matter density
perturbation $\delta$ in the conformal Newtonian gauge and its
fractional difference between two approximations,
$\left|\delta(\textrm{ours})-\delta(\textrm{GR})\right| /
\left[
\left|\delta(\textrm{ours})\right|+\left|\delta(\textrm{GR})\right|
\right]$. The gauge choice for presenting $\delta$ is made for
connecting to the late-time evolution of the matter density
perturbation that has been widely studied in the conformal
Newtonian gauge
\cite{Tsujikawa:2007gd,delaCruzDombriz:2008cp,Tsujikawa:2009ku}.


In addition, we present in these two figures two relevant
quantities: $c_{\chi}$, the fractional difference between
$\chi_\textrm{(approx)}$ and $\chi$, and $c_m \equiv (aH/k)^2
m$. The fractional difference $c_{\chi}$ gives a criterion for
the validity of our approximation, i.e., the smallness of it
indicates the validity of the approximation. The quantity $c_m$
is conventionally used to give a criterion for the validity of
the sub-horizon approximation in $f(R)$ gravity. One may use
$c_m$ to determine the starting time of invoking the late-time,
matter-dominated, sub-horizon approximate evolution equations.
This starting time will also be the ending time of invoking the
GR approximation if the initial condition of the late-time
evolution is given from solving the early-time evolution
equations with the GR approximation. While plotting
$\Theta_0(\textrm{GR})$ and $\delta(\textrm{GR})$ from an early
time $a=10^{-5}$ to a late time $a=0.1$, we plot
$\Theta_0(\textrm{ours})$ and $\delta(\textrm{ours})$ till the
time when $c_{\chi}=0.1$ (so as to the fractional difference),
before which our approximation is valid in assessing the effect
of the modification of gravity in $f(R)$ gravity.

Figure \ref{photon} shows that for the Fourier mode with
$k=0.1\,$Mpc$^{-1}$ the fractional difference in the CMB photon
density perturbation is about $1\%$ around the photon-baryon
decoupling time, $z_\textrm{dec}=1090$ ($a \sim 10^{-3}$), and
reaches as large as $10\%$ around $a=10^{-2}$. For
$k=0.1\,$Mpc$^{-1}$ the fractional difference is about one
order of magnitude smaller: $\lesssim 0.1\%$ around the
decoupling time; $\sim 1\%$ around $a=10^{-1.5} \simeq 0.03$.
This result indicates that the effect of the gravity
modification at early times in the $f(R)$ theory may not be
negligible compared to the accuracy of the CMB observations.
With regard to the matter density perturbation in Fig.\
\ref{deltam},  for $k=0.1\,$Mpc$^{-1}$ the fractional
difference is about $1\%$ around $a=10^{-2}$, which is
marginally negligible when compared to the current
observational accuracy, while for $k=0.01\,$Mpc$^{-1}$ it is
smaller: $\lesssim 10^{-3}$ before $a=10^{-1.5}\simeq 0.03$.



\clearpage

\clearpage


\section{Discussions} \label{sec:discussions}

In this paper we numerically solve the early-time evolution
equations of the linear cosmological perturbations in $f(R)$
gravity via an approximation we construct. With our
approximation we can fairly assess the effect of the gravity
modification in various $f(R)$ models on the early-time
evolution of the perturbations, thereby examining the validity
of the conventional GR approximation that neglects the
deviation from GR. In particular, we obtain the evolution of
the density perturbations of the CMB photons and matter, and
present the factional differences in these two quantities
between our approximation and the GR approximation. This
difference indicates the significance of the effect of gravity
modification on the evolution of the cosmological
perturbations.

We find that the effect of the gravity modification at early
times in $f(R)$ gravity may not be negligible, particularly for
the Fourier modes with shorter wavelengths such as
$k=0.1\,$Mpc$^{-1}$. Thus for self-consistency's sake, the GR
approximation is problematic, and a better treatment for the
early-time evolution is necessary, which our approximation may
provide. In our demonstration, even though the deviation from
GR looks tiny: $m \simeq -f_R  \sim 10^{-11}$ when $a \sim
10^{-3}$ and $\sim 10^{-8}$ when $a \sim 10^{-2}$, the
fractional difference in the CMB photon density perturbation
can reach $1\%$ at the photon-baryon decoupling time and even
$10\%$ around $a=10^{-2}$, which is significant compared to the
accuracy of the CMB observations. That is, even a tiny
deviation from GR at early times may induce a significant
effect on the cosmological perturbations.
This contradicts the conventional thinking. This situation is
analogous to the issue about the tight coupling between photons
and baryons before decoupling, which one is confronted with
when solving the evolution equations of the perturbations in
the standard cosmology \cite{Ma:1995ey}.

As a consequence, the CMB observations may provide a stringent
test to the currently viable $f(R)$ models, meanwhile giving
tighter constraints on $f(R)$ gravity than expected, and
further play an important role in distinguishing $f(R)$ gravity
from dark energy.
\begin{acknowledgments}
We thank the Dark Energy Working Group of the Leung Center for
Cosmology and Particle Astrophysics (LeCosPA), particularly
Huitzu Tu and Wolung Lee, for the helpful discussions.
Gu is supported by the Taiwan National Science Council (NSC)
under Project No.\ NSC98-2112-M-002-007-MY3,
Wang and Huang under NSC99-2112-M-002-009-MY3, and
Wu under NSC97-2112-M-002-026-MY3.
%
Chen is supported by the Taiwan NSC under Project No.\
NSC97-2112-M-002-026-MY3, by Taiwan's National Center for
Theoretical Sciences (NCTS), and by US Department of Energy
under Contract No.\ DE-AC03-76SF00515.
\end{acknowledgments}




\end{document}